\documentclass[11pt,a4]{article}
\usepackage[dvipdfmx]{graphicx}
\usepackage{cite}
\usepackage{color}
\usepackage{epstopdf}

\makeatletter
\def\slash#1{{\mathpalette\c@ncel{#1}}} 
\makeatother

\newcommand\beq{\begin{eqnarray}}
\newcommand\eeq{\end{eqnarray}}

\newcommand\la{\langle}
\newcommand\ra{\rangle}

\begin{document}

\title{\bf  Twist-3 effect from the longitudinally polarized
proton for \boldmath{$A_{LT}$} in hadron production from \boldmath{$pp$} collisions}

 \author{Yuji Koike$^{1}$, Daniel Pitonyak$^{2}$, and Shinsuke Yoshida$^{3}$
 \\[0.3cm]
{\normalsize\it $^1$Department of Physics, Niigata University, Ikarashi, Niigata 950-2181, Japan} \\[0.15cm]
{\normalsize\it $^2$RIKEN BNL Research Center, Brookhaven National Laboratory, Upton, NY 11973, USA} \\[0.15cm]
{\normalsize\it $^3$Key Laboratory of Quark and Lepton Physics (MOE) and Institute of Particle Physics,} \\
{\normalsize\it Central China Normal University, Wuhan 430079, China} \\[0.15cm]}

\date{\today}
\maketitle

\begin{abstract}
\noindent We compute the contribution from the longitudinally polarized proton to the twist-3 double-spin asymmetry $A_{LT}$
in inclusive (light) hadron 
production from proton-proton collisions,
i.e., $p^\uparrow \vec{p}\to h\,X$.  
We show that using the relevant QCD equation-of-motion relation and Lorentz invariance relation
allows one to eliminate the twist-3 quark-gluon correlator (associated with the longitudinally polarized proton) in favor of one-variable twist-3 quark distributions and the (twist-2) transversity parton density.  
Including this result with the twist-3 pieces associated with 
the transversely polarized proton and
unpolarized final-state hadron (which have already been calculated in the literature), we now have the complete leading-order cross section for this process.  
\end{abstract}

\newpage
\section{Introduction}

Twist-3 observables in high-energy semi-inclusive reactions
provide us with an important opportunity
to test theoretical frameworks for QCD hard processes and to
understand the quark-gluon substructure of hadrons beyond the conventional parton model.  
Well-known examples are the experimental observation of hyperons with
large transverse polarization
produced in unpolarized proton-proton collision, $pp\to \Lambda^\uparrow X$~\cite{Bunce:1976yb,Smith87,Ho90_1,Ho90_2,Ho90_3},
and the transverse single-spin (or left-right) asymmetry (SSA) $A_N$ of a produced hadron
in the collision between a transversely polarized proton and an unpolarized proton,
$p^\uparrow p\to h\,X$ ($h=\pi,K,\eta$, 
etc.)~\cite{Klem:1976ui,Adams:1991rw,Krueger:1998hz,
Allgower:2002qi,Adams:2003fx,Adler:2005in,Lee:2007zzh,:2008mi,Adamczyk:2012qj,Bland:2013pkt,Adare:2013ekj}.
The magnitude of the asymmetries were as large as a few tens of percent in the
forward direction.  
In collinear factorization, these SSAs appear as twist-3 observables.  They are driven by
multi-parton (quark-gluon or purely gluonic)
correlations~\cite{Efremov:1981sh,Qiu:1991pp} 
either in the initial-state hadrons or in the final-state fragmentation process.   
The formalism for deriving the twist-3 cross section for SSAs has been 
well developed, and  the formulae
involve the relevant multi-parton correlation functions instead of the usual (twist-2) parton densities or fragmentation
functions~\cite{Kanazawa:2000hz,
Kouvaris:2006zy,Eguchi:2006qz,Eguchi:2006mc,Zhou:2008,Yuan:2009dw,
Kang:2010zzb,Metz:2012ct,
Kanazawa:2013uia,Beppu:2010qn,Koike:2006qv,Koike:2007rq,Koike:2009ge,Metz:2010xs,
Beppu:2013uda,Kanazawa:2014nea,Koike:2015zya}.  
The $A_N$ data for $\pi$, $K$, $\eta$, and jet 
production obtained 
at the Relativistic Heavy Ion Collider (RHIC)
have been analyzed using this formalism~\cite{Kouvaris:2006zy,Kanazawa:2010au,Gamberg:2013kla,Kanazawa:2014dca}.\footnote{Data from RHIC is on tape for $A_N$ in prompt photon production and several predictions exist for this asymmetry within collinear factorization\cite{Gamberg:2012iq,Kanazawa:2014nea,Gamberg:2013kla}.}

Besides these large SSAs, the double-spin asymmetry (DSA) $A_{LT}$ for particle production (direct photon, Drell-Yan lepton pair, 
hadron, jet, etc.)
in collisions between longitudinally and transversely polarized protons, $p^\uparrow \vec{p}\to C\,X$, 
is also a twist-3 
observable~\cite{Jaffe:1991kp,Koike:2008du,Liang:2012rb,Metz:2012fq,Hatta:2013wsa,Koike:2015yza}.\footnote{$A_{LT}$ in $ep$ collisions
is also an interesting twist-3 asymmetry and has been studied in Refs.~\cite{Kang:2011jw,Kanazawa:2014tda,Kanazawa:2015ajw}.}    
Unlike SSAs, which are naively ``T-odd" effects, DSAs like $A_{LT}$ are naively ``T-even," which
leads inherently to different forms for the corresponding twist-3 cross section (see the discussion below Eq.~(\ref{twist3})). 
Therefore, $A_{LT}$ and 
$A_N$ probe different yet complimentary aspects of hadronic structure, and both are critical to test the underlying mechanism for these asymmetries.  Surprisingly, RHIC has never run an experiment for $A_{LT}$ despite being the only facility in the world with polarized proton beams and having measured every other combination of proton spins ($A_N$, $A_L$, $A_{TT}$, $A_{LL}$). 

In this paper we compute the polarized cross section 
for $A_{LT}$ in the production of an unpolarized (light) hadron $h$ from proton-proton collisions,
\beq
p(P,S_{\perp})+p(P',{\Lambda})\to h(P_h)+X\,,
\label{pphX}
\eeq
where $S_\perp$ is the transverse spin vector for the nucleon $A$,
${\Lambda}$ is the helicity of the longitudinally polarized nucleon $B$, and
the momenta of the particles are shown.  
In the framework of collinear factorization, 
the first nonvanishing contribution to the cross section appears at twist-3, and it
receives three contributions,
\begin{eqnarray} 
d\sigma(P_{h},S_{\perp},\Lambda) &=& \,H\otimes f_{a/A(3)}\otimes 
f_{b/B(2)}\otimes D_{h/c(2)} \nonumber \\[-0.2cm]
&&+ \,H'\otimes f_{a/A(2)}\otimes f_{b/B(3)}\otimes D_{h/c(2)} \nonumber \\[-0.2cm]
&&+ \,H''\otimes f_{a/A(2)}\otimes f_{b/B(2)}\otimes D_{h/c(3)}\,,
\label{twist3}
\end{eqnarray} 
where $f_{a/A(3)}$ 
represents the twist-3 distribution function
for parton species $a$ ($a=q ,\ \bar{q},\ g$) 
in nucleon $A$ 
with the subscript $(3)$ indicating the twist (and similar for $f_{b/B(3)}$).  
Likewise, $D_{h/c(3)}$ represents
the twist-3 fragmentation function for the parton species $c$ into the final-state hadron $h$.  
The factors $H$, $H'$, and $H''$ are the partonic hard cross sections for each contribution, and 
$\otimes$ represents a convolution in the appropriate momentum fractions.  

So far, the leading-order (LO) cross section was derived for
the first term~\cite{Metz:2012fq} and the third term~\cite{Koike:2015yza} in Eq.~(\ref{twist3}).  
The first line of (\ref{twist3}) involves twist-3 distributions in the transversely polarized nucleon
coupled to the twist-2 helicity distribution.  
Unlike the SSA for $p^\uparrow p\to h\,X$, 
the partonic hard part for this term is given 
as a non-pole contribution~\cite{Liang:2012rb,Metz:2012fq}.  
In the third line of (\ref{twist3}), the real part of the unpolarized chiral-odd twist-3 quark-gluon fragmentation function couples to the transversity parton density~\cite{Koike:2015yza}.  This is in contrast to SSAs, where
the {\it imaginary part} of the same quark-gluon twist-3 fragmentation function contributes~\cite{Metz:2012ct,Kanazawa:2013uia}.  
A recent analysis suggests that this imaginary part can be the main cause of the large $A_N$ observed for pion production
in $pp$ collisions at RHIC~\cite{Kanazawa:2014dca}.  This new insight is what motivated the calculation of the third line in Eq.~(\ref{twist3}) for the $A_{LT}$ case~\cite{Koike:2015yza}.
Again we emphasize that $A_{LT}$ in $p^\uparrow \vec{p}\to h\,X$ is a unique
quantity that should be measured at RHIC. 

To complete the LO cross section for the process (\ref{pphX}), we will compute
the second term in Eq.~(\ref{twist3}), where, as we will see in Sec.~3, chiral-odd twist-3 distributions for the longitudinally polarized nucleon
enter along with the transversity parton density (the latter shows up when one employs QCD equation-of-motion and Lorentz invariance relations).  Both of these couple to the transversity function for the transversely polarized nucleon.
We note that two twist-3 terms analogous to the first two lines in Eq.~(\ref{twist3}) 
(with the fragmentation functions omitted)
contribute to $A_{LT}$ in Drell-Yan when one integrates
over the transverse momenta of the lepton pair, and both pieces are of a similar magnitude~\cite{Koike:2008du}.  
Therefore, it is possible that the second term of (\ref{twist3}) for hadron production is just as important as the first and
brings a non-negligible contribution.  In addition, as alluded to above, the third term might also be significant (as in $A_N$).  Thus, a detailed numerical study of all three parts of $A_{LT}$ will be needed and is the subject of future work.

The rest of this paper is organized as follows:
in Sec.~2 we summarize the twist-3 distribution functions in the nucleon relevant for this computation
and the relations among them.  
In Sec.~3, we derive the LO cross section for the second term of Eq.~(\ref{twist3}).  
We will see that, owing to a simple form of the partonic hard cross sections, the effect of the twist-3 quark-gluon correlation function 
in the longitudinally polarized nucleon can be
expressed in terms of one-variable
twist-3 quark distributions and the transversity  parton density.  
Sec.~4 is devoted to a brief summary.


\section{Twist-3 distribution functions for a longitudinally polarized proton \label{s:Def}}

In this section we summarize the distribution functions in the nucleon 
relevant to our study.
We first have a quark correlator in the nucleon that gives two chiral-odd polarized functions needed in our calculation~\cite{Jaffe:1991kp},
\beq
M^q_{ij}(x)&=&\int{d\lambda\over 2\pi}e^{i{\lambda}x}\la PS|\bar{\psi}_j(0)\psi_i(\lambda n)|PS\ra \nonumber\\
&=&{1\over 2}(\gamma_5\slash{S}_{\perp}\slash{p})_{ij}h^q_1(x)+{M_N\over 2}
\Lambda (i\gamma_5\sigma^{np})_{ij}h^q_L(x)+\cdots,
\label{M}
\eeq
where
$\psi_i$ is a quark field with spinor index $i$, $M_N$ is the nucleon mass, 
$S$ is the nucleon spin vector normalized as $S^2=-1$, and $\Lambda=M_N\left(S\cdot n\right)$ is
its helicity.
We also introduced two lightlike vectors $p^\mu$ and
$n^{\mu}$, where $P=p+(M_N^2/2)n$ and $p\cdot n=1$
with the only nonzero components $p^+=P^+$ and $n^-$ for the nucleon moving in the
$+z$-direction. 
For simplicity, here and below we suppress the gauge-link operators
and use the shorthand
$\sigma^{np}\equiv \sigma^{\alpha\beta}n_{\alpha}p_{\beta}$.  
The $F$-type twist-3 distribution in the longitudinally polarized proton is defined as~\cite{Kodaira:1999}
\beq
M^{q,\alpha}_{F\,ij}(x_1,x_2)&=&\int{d\lambda\over 2\pi}\int{d\mu\over 2\pi}
e^{i{\lambda}x_1}e^{i\mu(x_2-x_1)}\la PS|\bar{\psi}_j(0)gF^{\alpha n}(\mu n)\psi_i(\lambda n)|PS\ra \nonumber\\
&=&i{M_N\over 2}g_{\perp}^{\alpha\beta}\Lambda(\gamma_5\gamma_{\beta}\slash{p})_{ij}H^q_{FL}(x_1,x_2)+\cdots, \label{F-type}
\eeq
where $F^{\alpha n}$ is the gluon field strength tensor and 
$g_{\perp}^{\alpha\beta}\equiv g^{\alpha\beta}-p^{\alpha}n^{\beta}-p^{\beta}n^{\alpha}$. 
From Hermiticity and $PT$-invariance, 
$H_{FL}(x_1,x_2)$ is shown to be real and satisfies the 
symmetry property 
\beq
H^q_{FL}(x_1,x_2)=-H^q_{FL}(x_2,x_1).
\eeq
The $D$-type twist-3 distribution
$H_{DL}(x_1,x_2)$ is defined 
by the replacement $gF^{\alpha w}(\mu n)\rightarrow 
D^{\alpha}({\mu n})=\partial^{\alpha}-igA^{\alpha}(\mu n)$ in (\ref{F-type}), and is related to
$H^q_{FL}(x_1,x_2)$ as
\beq
H^q_{DL}(x_1,x_2)={\cal P}{1\over x_1-x_2}\,H^q_{FL}(x_1,x_2)+\delta(x_1-x_2)\tilde{h}^q_{L}(x_2),
\label{GIR}
\eeq
where ${\cal P}$ indicates the
principal value.  The function $\tilde{h}_{L}(x)$ is another real twist-3 distribution function, which is defined as
\beq
&&\hspace{-0.6cm}M^{q,\alpha}_{\partial\,ij}(z)
=\!\!\lim_{z_\perp\to 0}\!\int \!\!{d\lambda \over 2\pi} e^{i\lambda x} {\partial \over \partial z_{\perp\alpha} }
\la PS|\bar{\psi}_j(0)[0,\infty n][\infty n, \infty n +z_\perp][\infty n+ z_\perp,\lambda n +z_\perp]
\psi_i(\lambda n+z_\perp)|PS\ra \nonumber\\
&=&\int{d\lambda\over 2\pi}e^{i\lambda x}
\la PS|\bar{\psi}_j(0)D^{\alpha}(\lambda n)\psi_i(\lambda n)|PS\ra 
+\int{d\lambda\over 2\pi}e^{i\lambda x}\int_\lambda^\infty d\mu
\la PS|\bar{\psi}_j(0)igF^{\alpha n}(\mu n)\psi_i(\lambda n)|PS\ra \nonumber\\
&=&i{M_N\over 2}g_{\perp}^{\alpha\beta}\Lambda(\gamma_5\gamma_{\beta}\slash{p})_{ij}\tilde{h}^q_L(x)+\cdots,
\label{Mpartial}
\eeq
where in the first line we explicitly wrote the gauge links $[\infty n+ z_\perp,\lambda n +z_\perp]$, etc., so that
the meaning of the derivative becomes clear.  
Using the QCD equation-of-motion, $h_L(x)$ can be expressed in terms of $H_{FL}(x_1,x)$ 
and $\tilde{h}_L(x)$
as
\beq
h^q_L(x)&=&-{1\over x}\int^1_{-1} dx_1\,\left(H^q_{DL}(x_1,x)+H^q_{DL}(x,x_1)\right) \nonumber\\
&=&-{2\over x}\int_{-1}^1 dx_1\,{\cal P}{1\over x_1-x}\,H^q_{FL}(x_1,x)-{2\over x}\tilde{h}^q_L(x). 
\label{EOM}
\eeq
In addition, the operator product expansion gives another relation 
among $h_L(x)$, $h_1(x)$, and $H_{FL}(x_1,x_2)$ as~\cite{Kodaira:1999}
\beq
-x^2{d\over dx}\left({1\over x}h^q_L(x)\right)
=2h^q_1(x)+2\int_{-1}^1 dx_1\,{\cal P}{1\over x-x_1}\left({\partial\over \partial x}-
{\partial\over \partial x_1}\right)H^q_{FL}(x,x_1).\label{OPE}
\eeq
The combination of (\ref{EOM}) and (\ref{OPE}) leads to
\beq
{d\tilde{h}^q_L(x)\over dx}-h^q_1(x)+h^q_L(x)=
2\int_{-1}^1 dx_1\,{\cal P}{1\over (x-x_1)^2}H^q_{FL}(x,x_1),\label{LIR}
\eeq
which is known as a Lorentz invariance relation in the literature~\cite{Kanazawa:2015ajw}.  
In Sec.~3, we will see the relations (\ref{EOM}) and (\ref{LIR}) lead to
a simple form for the cross section for the second term of (\ref{twist3}).


\section{Calculation of the polarized cross section for \boldmath{$A_{LT}$}\label{s:Def}}

\begin{figure}[t]
\begin{center}
\hspace{-1cm}
\includegraphics[width=5cm]{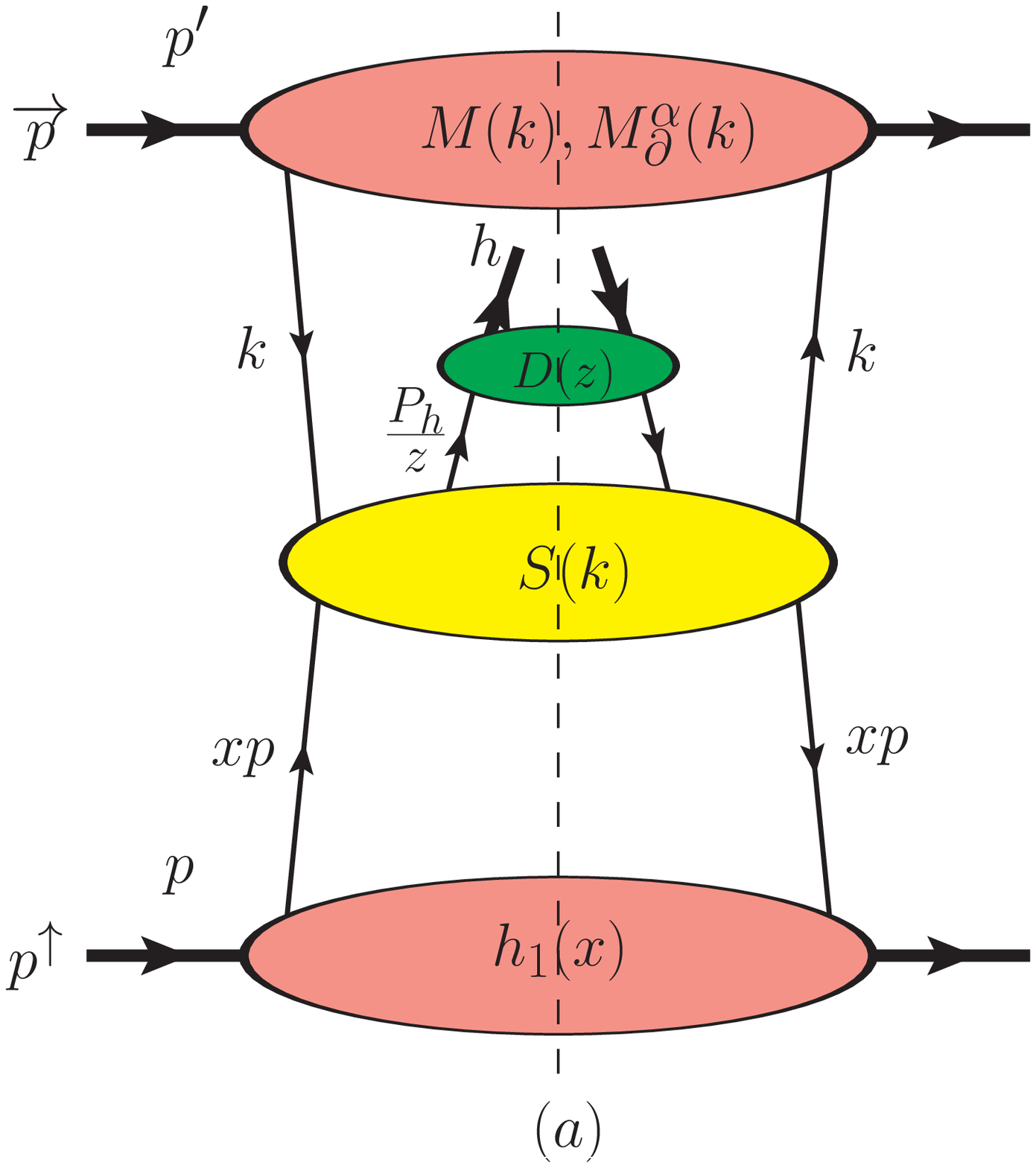}\hspace{0.5cm}
\includegraphics[width=5cm]{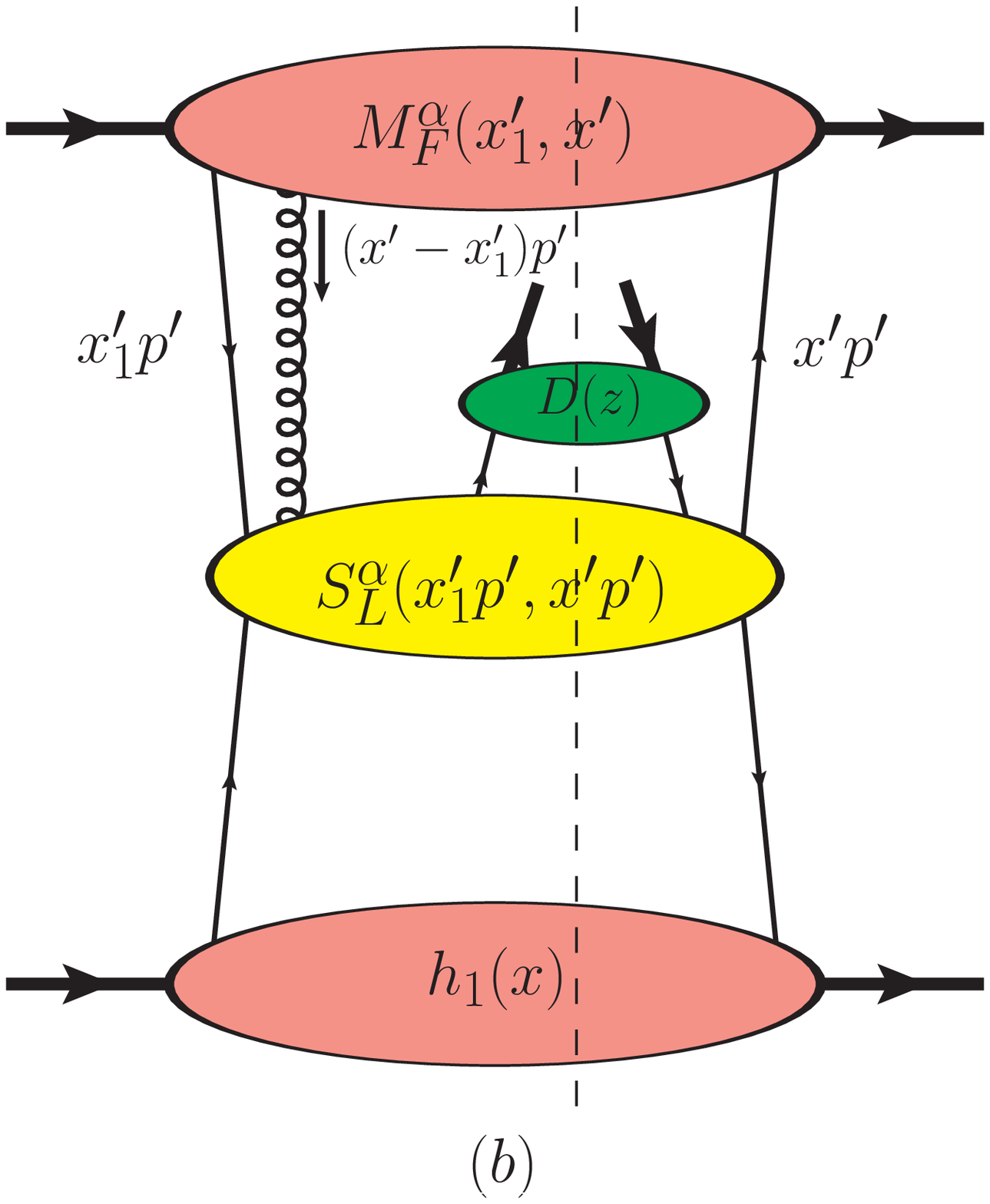}\hspace{0.5cm}
\includegraphics[width=5cm]{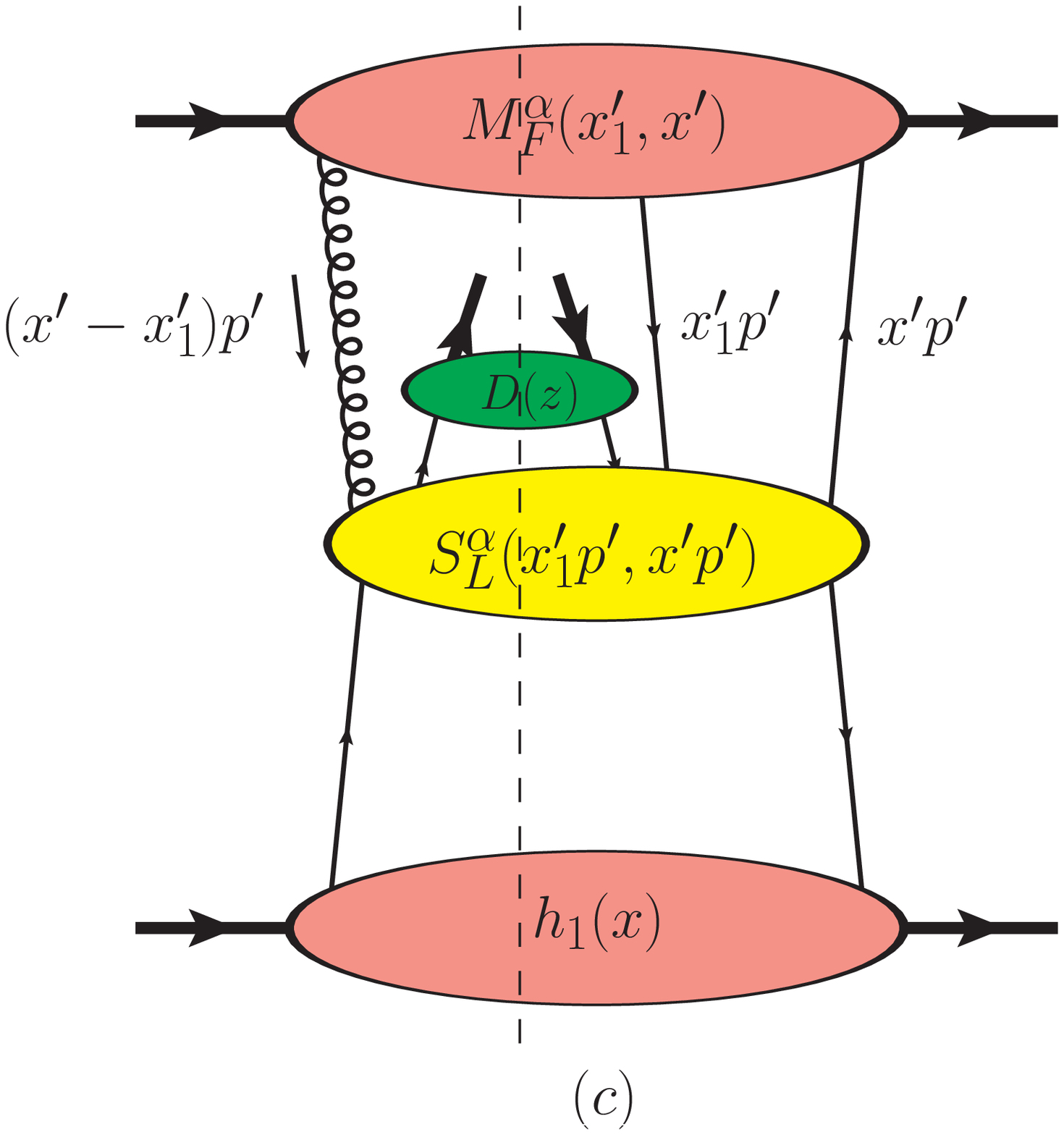}
\caption[]{Generic diagrams for the contribution to the process (\ref{pphX}) from the second term in Eq.~(\ref{twist3}).
The correlators for the longitudinally polarized nucleon (upper blob) couple to the transversity
distribution (lower blob).  Diagram (a) gives rise to the first and second terms
in (\ref{Xsec0}), and (b) and (c) are for the third term in (\ref{Xsec0}). 
Mirror diagrams of (b) and (c) also contribute, which are included in Eq.~(\ref{Xsec0}). }
\label{fig1}
\end{center}
\end{figure}

We now derive the cross section for the second term of Eq.~(\ref{twist3}).  
As mentioned before, the twist-3 cross section for the naively T-even $A_{LT}$
arises from non-pole contributions.   
The method of the calculation 
has been formulated
both in Feynman gauge~\cite{Kanazawa:2013uia,Hatta:2013wsa} and 
lightcone gauge~\cite{Liang:2012rb,Metz:2012fq,Metz:2012ct}, 
and it has been confirmed that
they give identical results for the twist-3 cross section in terms of the gauge-invariant 
distribution and fragmentation functions defined in the previous 
section~\cite{Koike:2015yza,Kanazawa:2014tda,Kanazawa:2015jxa}.  
Here we follow the Feynman gauge formulation (but have checked that the same result is achieved in lightcone gauge), which has an advantage that the gauge invariant correlation
functions appear manifestly.  
Since we are interested in the twist-3 effect from the longitudinally polarized nucleon,
we factorize the transversity distribution $h_1(x)$ and the unpolarized fragmentation function
for the hadron $D(z)$ from the rest of the cross section and perform a collinear
expansion of the hard part.
The generic diagrams for this contribution is shown in Fig.~1.
According to the general formalism developed in \cite{Kanazawa:2013uia}, 
the twist-3 cross section is obtained as
\beq
&&\hspace{-0.5cm}E_{h}{d\sigma(S_{\perp},\Lambda)\over d^3P_h} \nonumber\\
&=&{1\over 16\pi^2S}\int{dx\over x}h_1(x)\int{dz\over z^2}D(z)\left\{
\int dx'\,{\rm Tr}\left[M(x')S(x'p')\right] +i\omega^{\alpha}_{\ \beta}\int dx'\,{\rm Tr}\left[M_{\partial}^{\beta}(x')
\left.{\partial S(k) \over \partial k^{\alpha}}\right|_{k=x'p'}\right] \right.\nonumber\\
&&\qquad\left. 
+2i\omega^{\alpha}_{\ \beta}\int dx'\int dx'_1\,{\cal P}{1\over x'_1-x'}
{\rm Tr}\left[M_F^\beta(x'_1,x')S_{L\alpha}(x'_1p',x'p')\right]
\right\},
\label{Xsec0}
\eeq
where $S=(P+P')^2$ is the center-of-mass energy squared, 
$M(x')$, $M_\partial^\beta (x')$, and $M^\beta_F(x_1',x')$ are, respectively, defined in
Eqs.~(\ref{M}), (\ref{Mpartial}), and (\ref{F-type})  
with $p$ and $n$ replaced by
$p'$ and $n'$ (similarly defined for the momentum $P'$ by $P'=p'+(M_N^2/2)n'$ and $p'\cdot n'=1$), and
$\omega^\alpha_{\ \beta}=g^\alpha_{\ \beta} -p'^\alpha n'_\beta$. 
The partonic hard parts $S(k)$ 
and $S_{L\alpha}(x'_1p',x'p')$ are shown by the middle blobs of Fig.~1(a) and Fig.~1(b),(c), respectively.  (It is understood that $S$ and $S_{L\alpha}$ also depend on $xp$ and $P_h/z$.)
Here $S_{L\alpha}(x'_1p',x'p')$ represents the hard part for the diagram in which
the coherent gluon line from $M_F^\beta(x'_1,x')$ is located in the left of the cut,
and the effect of the mirror diagrams is taken into account by the principal value prescription
and the factor of 2 in the third term of Eq.~(\ref{Xsec0}).
The LO diagrams for the hard parts are shown in Figs.~2--4:
they correspond to the
$qq\to qq$ channel\footnote{Here $ab\to cd$ implies that parton $a$
is from $p^\uparrow$, $b$ is from $\vec{p}$, and $c$ fragments into the hadron $h$.} (Fig. 2), 
$\bar{q}q\to q'\bar{q}'$, $\bar{q}q\to \bar{q}'q'$, $\bar{q}q\to q\bar{q}$, 
$\bar{q}q\to \bar{q}q$ channels (Fig. 3),
and $\bar{q}q\to gg$ channel (Fig. 4).
Inspecting these diagrams, it is not difficult to find that
$S_{L\alpha}(x'_1p',x'p')$ depends on $x_1'$ only through the factor $1/(x_1'-x')$ and $1/x_1'$.  
Therefore the cross section 
can be decomposed as
\beq
&&E_{h}{d\sigma(S_{\perp},\Lambda)\over d^3P_h} 
={2\alpha_s^2M_N\Lambda\over S}(S_{\perp}\cdot P_h)\sum_i\sum_{a,b,c}
\int_0^1{dx\over x}h^a_1(x)\int_0^1{dz\over z^3}D^c(z)\int_0^1 dx'\delta(\hat{s}+\hat{t}+\hat{u})\nonumber\\
&&\qquad\qquad\times\left[
h^b_L(x')\hat{\sigma}^i_L+{\tilde{h}^b_L(x')\over x'}\hat{\sigma}^i_{ND}+{d\tilde{h}^b_L(x')\over dx'}\hat{\sigma}^i_{D} 
\right.\nonumber\\
&&\qquad\qquad\left.+{1\over x'}\int_{-1}^1 dx'_1\,{\cal P}{1\over x'_1-x'}H^b_{FL}(x'_1,x')\hat{\sigma}^i_{F1}
+2\int_{-1}^1 dx'_1\,{\cal P}{1\over (x'_1-x')^2}H^b_{FL}(x'_1,x')\hat{\sigma}^i_{F2}\right. \nonumber\\
&&\qquad\qquad\left.+\int_{-1}^1 dx'_1\,{\cal P}{1\over x'_1(x'_1-x')}H^b_{FL}(x'_1,x')\hat{\sigma}^i_{SFP}
\right],  
\label{Xsec1}
\eeq
where $\sum_i\sum_{a,b,c}$ indicates a sum over channels $i$ and parton flavors in each channel (where $\{a,b\}\!\in\! \{q,\bar{q}\}$, $c\in\! \{q,\bar{q},g\}$). The partonic hard cross sections
$\hat{\sigma}_L$, $\hat{\sigma}_{ND}$, $\hat{\sigma}_D$, $\hat{\sigma}_{F1}$, $\hat{\sigma}_{F2}$, $\hat{\sigma}_{SFP}$
are independent of $x'_1$ and are functions of
the Mandelstam variables 
\beq
\hat{s}=\left(xp+x'p'\right)^2,\hspace{5mm}\hat{t}=\left(xp-{P_h\over z}\right)^2,\hspace{5mm}\hat{u}=\left(x'p'-{P_h\over z}\right)^2.
\eeq
By extracting the $1/x'_1$ component of $S_{L\alpha}(x'_1p',x'p')$ we can see that
$\hat{\sigma}_{SFP}$ has a structure identical to a SSA soft-fermion-pole (SFP) cross section (besides the projection tensor) 
with $x'_1=0$~\cite{Koike:2009ge,Kanazawa:2014nea,Koike:2015zya}.  
By direct computation of all channels,
we find that $\hat{\sigma}_{SFP}=0$, $\hat{\sigma}_{ND}=\hat{\sigma}_{F1}$,
and the contribution from Fig.~1(c) is identically zero.  
This vanishing $\hat{\sigma}_{SFP}$ is reminiscent of the fact that
the SFP hard parts of the chiral-odd contribution
to $pp\to\Lambda^\uparrow X$ and $p^\uparrow p\to \gamma X$ (i.e., the piece involving twist-3 distributions for the unpolarized proton) 
vanish~\cite{Kanazawa:2014nea,Koike:2015zya}. 
Accordingly, 
using Eqs.~(\ref{EOM}) and (\ref{LIR}) in Eq.~(\ref{Xsec1}), one can eliminate
$H_{FL}(x_1',x')$ in favor of $h_{1}(x')$, $h_L(x')$, and $\tilde{h}_L(x')$
and obtain the twist-3 cross section as
\beq
E_{h}{d\sigma(S_{\perp},\Lambda)\over d^3P_h}
&=&{2\alpha_s^2M_N\Lambda\over S}(S_{\perp}\cdot P_h)\sum_i\sum_{a,b,c}\int_0^1{dx\over x}h^a_1(x)\int_0^1{dz\over z^3}
D^c(z)\int_0^1 dx'\,\delta(\hat{s}+\hat{t}+\hat{u}) \nonumber\\[-0.3cm]
&&\qquad\qquad
\times\left[h^b_1(x')\hat{\sigma}^i_{1}+h^b_L(x')\hat{\sigma}^i_{2}
+{d \tilde{h}^b_L(x') \over dx'}\hat{\sigma}^i_{3}
\right],
\label{final}
\eeq
with
\begin{equation}
\hat{\sigma}_1 \equiv \hat{\sigma}_{F2}\,,\quad \hat{\sigma}_{2}\equiv \hat{\sigma}_L-\hat{\sigma}_{F2}-{1\over 2}\hat{\sigma}_{F1}\,,\quad \hat{\sigma}_{3}\equiv \hat{\sigma}_{D}-\hat{\sigma}_{F2}.  
\end{equation}
The partonic cross section for each channel reads\footnote{$N=3$ is the number of colors and $C_F=(N^2-1)/2N=4/3$.}
\begin{figure}[t]
\begin{center}
\includegraphics[width=9cm]{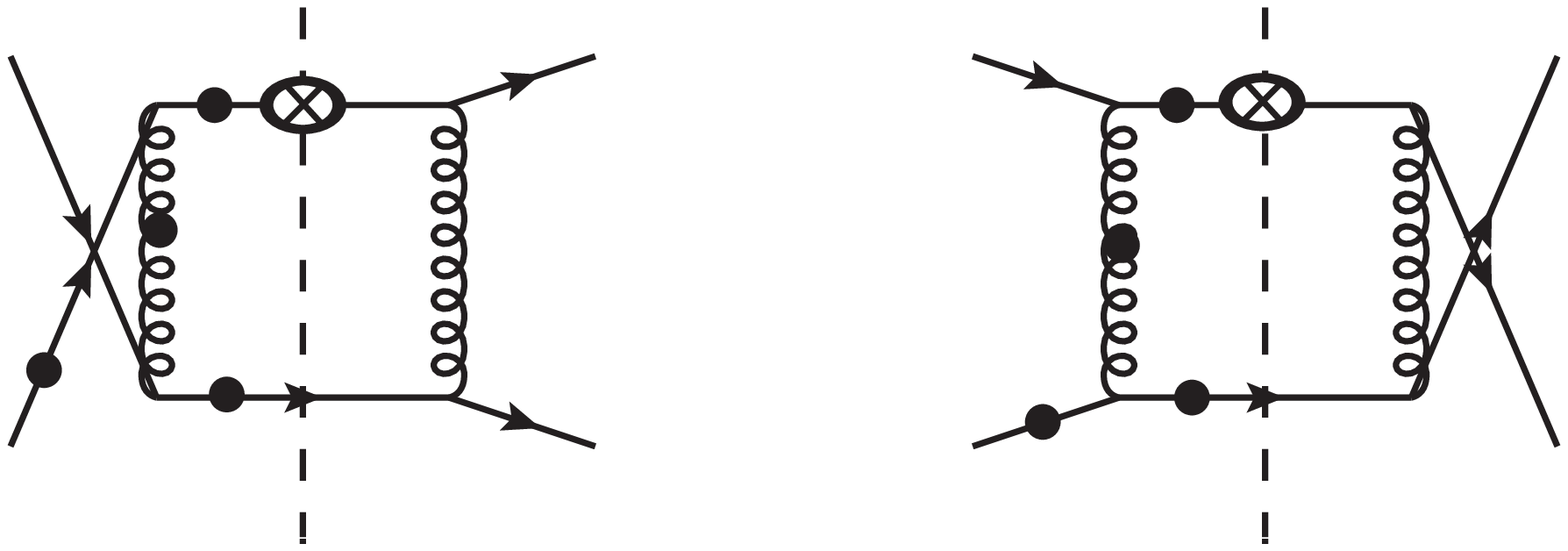} \\
\includegraphics[width=14cm]{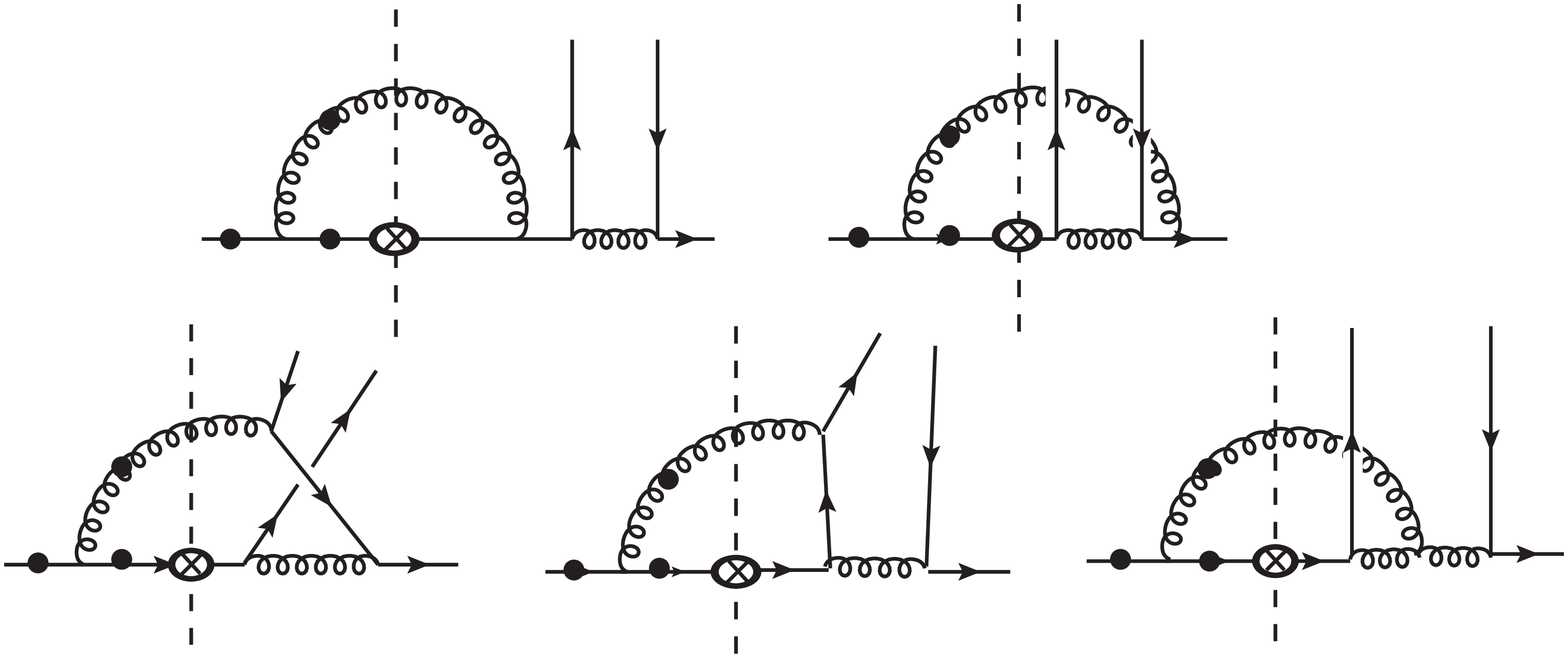}
\vspace{-0.3cm}
\caption[]{Feynman diagrams in
the $qq\to qq$ channel for the partonic hard parts $S(k)$ and $S_{L\alpha}(x'_1p',x'p')$ in (\ref{Xsec0}). 
Only the top two diagrams contribute to $S(k)$, while all the diagrams contribute to $S_{L\alpha}(x'_1p',x'p')$.  
The circled cross indicates 
the fragmentation insertion.  For $S_{L\alpha}(x'_1p',x'p')$, it is understood for each diagram that
the coherent gluon line coming out of the
longitudinally polarized nucleon matrix element (upper side) attaches to one of the dots.   
Mirror diagrams also contribute, 
which is taken into account in (\ref{Xsec0}).}
 \label{f:qqqq}
\end{center}
\end{figure}
\begin{figure}[t]
\begin{center}
\includegraphics[width=14cm]{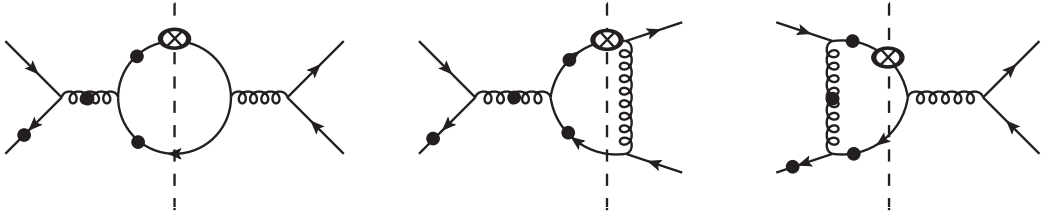}
\vspace{-0.3cm}
\caption[]{The same as Fig. 2, but for the
$\bar{q}q\to q'\bar{q}'$, $\bar{q}q\to \bar{q}'q'$, $\bar{q}q\to q\bar{q}$, 
$\bar{q}q\to \bar{q}q$ channels.  
Only the first diagram contributes in the $\bar{q}q\to q'\bar{q}'$ and $\bar{q}q\to \bar{q}'q'$ channels.  }
 \label{f:qqbarqqbar}
\end{center}
\end{figure}
\begin{figure}[t]
\begin{center}
\includegraphics[width=14cm]{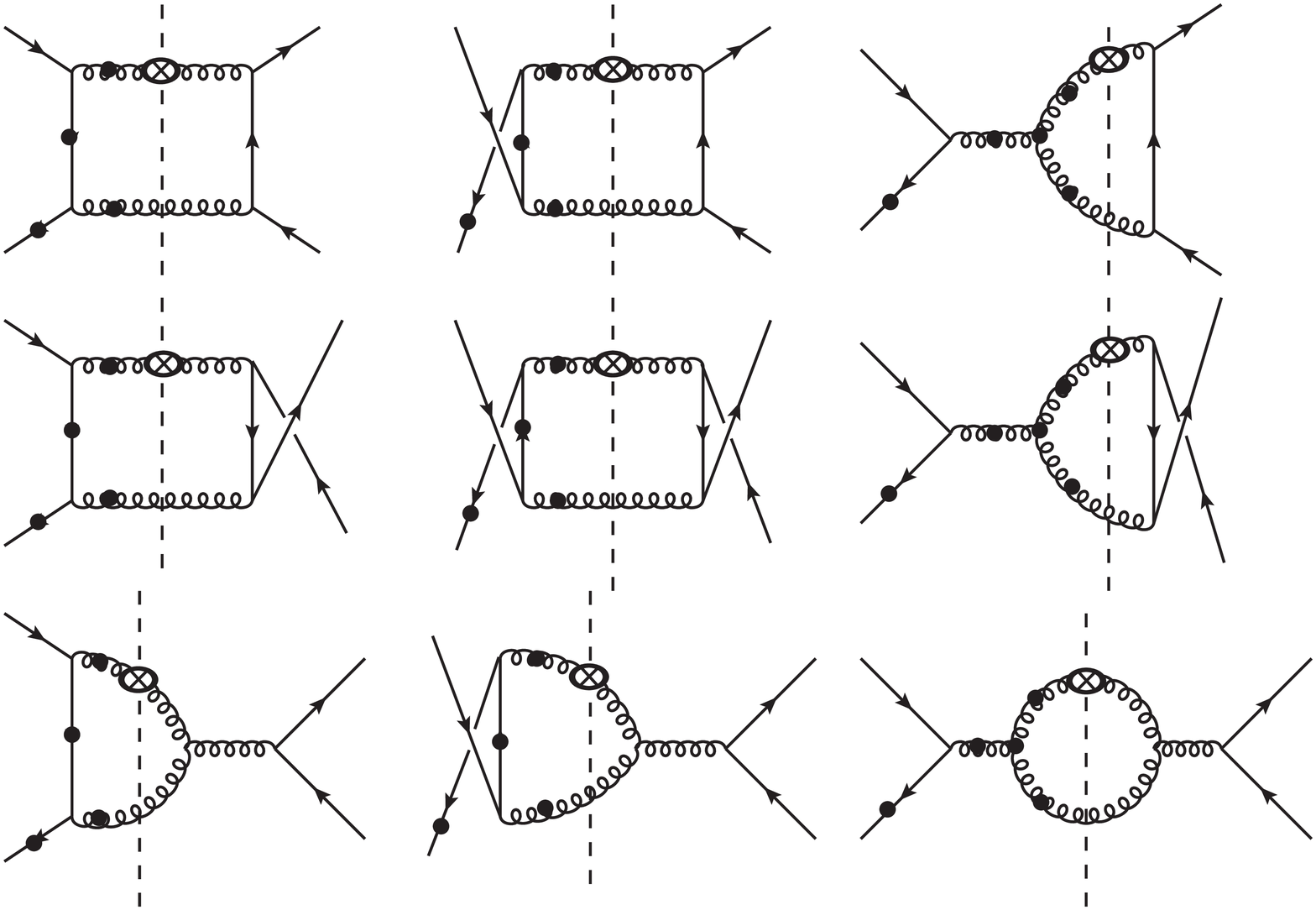}\\
\includegraphics[width=14cm]{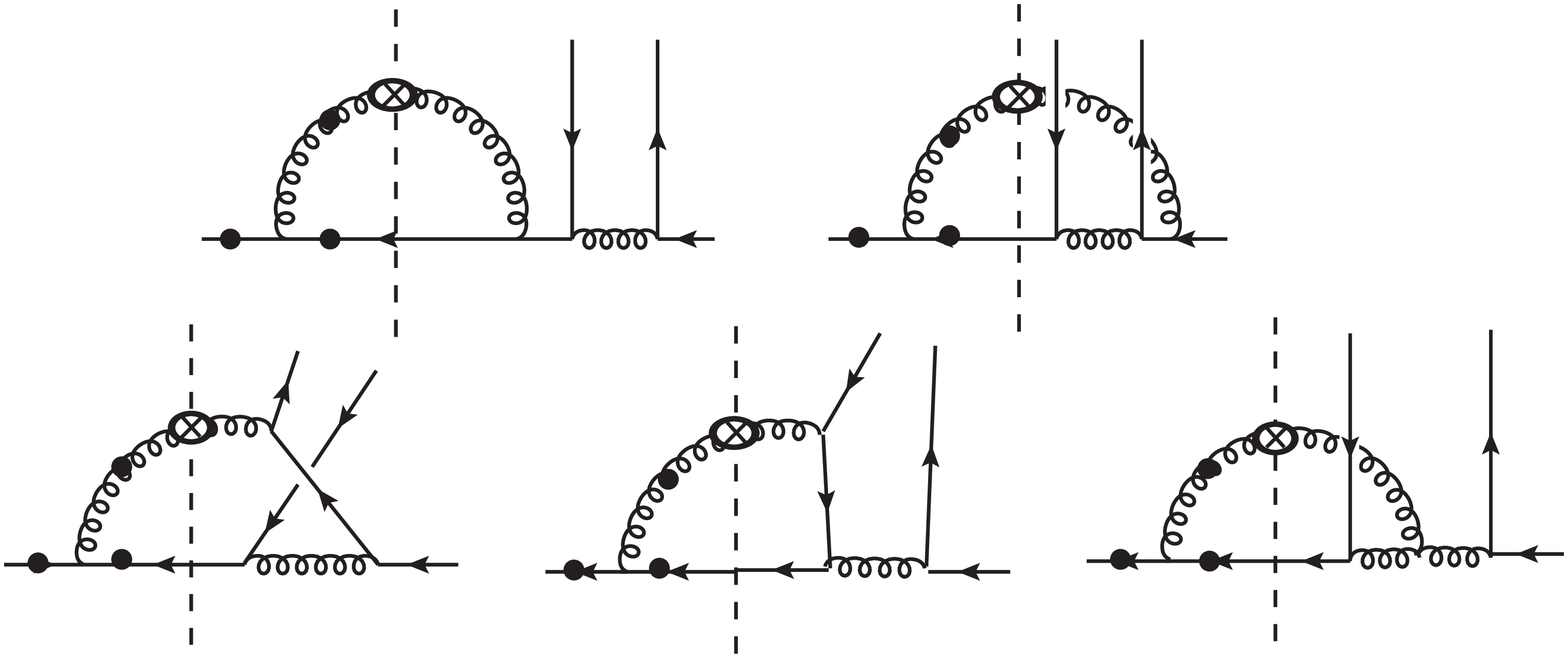}
\vspace{-0.3cm}
\caption[]{The same as Fig. 2, but for the $\bar{q}q\to gg$ channel.
Only the top nine diagrams contribute to $S(k)$, while all the diagrams contribute to $S_{L\alpha}(x'_1p',x'p')$. }
 \label{f:qbarqgg}
\end{center}
\end{figure}
\vspace{2mm}

\noindent
(i) $qq\to qq$ channel:
\beq \label{e:sigma1}
\hat{\sigma}_{1}=-{1\over N^3}{\hat{t}-\hat{u}\over \hat{t}\hat{u}},\qquad
\hat{\sigma}_{2}=\left({1\over N}+{1\over N^3}\right){\hat{t}-\hat{u}\over 2\hat{t}\hat{u}}, 
\qquad 
\hat{\sigma}_{3}=
-{1\over N}{1\over \hat{t}}+{1\over N^3}{1\over \hat{u}}.
\eeq
(ii) $\bar{q}q\to q'\bar{q}'$ channel:
\beq
\hat{\sigma}_{1}={\hat{t}\over \hat{s}^2}+{1\over N^2}{2\over \hat{s}},\qquad
\hat{\sigma}_{2}=-{\hat{u}\over \hat{s}^2}+{1\over N^2}{2\hat{u}-\hat{s}\over 
\hat{s}^2},\qquad
\hat{\sigma}_{3}={1\over \hat{s}}+{1\over N^2}{\hat{u}-2\hat{s}\over \hat{s}^2}.
\eeq
(iii) $\bar{q}q\to \bar{q}'q'$ channel:
\beq
\hat{\sigma}_{1}=-{\hat{u}\over \hat{s}^2}-{1\over N^2}{2\over \hat{s}},\qquad
\hat{\sigma}_{2}={\hat{t}\over \hat{s}^2}+{1\over N^2}
{\hat{s}-2\hat{t}\over \hat{s}^2}, \qquad
\hat{\sigma}_{3}={1\over N^2}{2\hat{s}+\hat{u}\over \hat{s}^2}.
\eeq
(iv) $\bar{q}q\to q\bar{q}$ channel:
\beq
\hat{\sigma}_{1}&=& {\hat{t}\over \hat{s}^2}+{1\over N^2}{2\over \hat{s}} + {1\over N}{1\over \hat{s}}-{1\over N^3}{1\over \hat{u}},\nonumber\\
\hat{\sigma}_{2}&=&-{\hat{u}\over \hat{s}^2}+{1\over N^2}
{2\hat{u}-\hat{s}\over \hat{s}^2}-{1\over N}{\hat{t}\over 2\hat{s}\hat{u}}-{1\over N^3}{\hat{t}+4\hat{u}\over 
2\hat{s}\hat{u}},\nonumber \\
\hat{\sigma}_{3}&=&{1\over \hat{s}}+{1\over N^2}{\hat{u}-2\hat{s}\over \hat{s}^2}-{1\over N^3}{\hat{u}-\hat{s}\over \hat{s}\hat{u}}.
\eeq
(v) $\bar{q}q\to \bar{q}q$ channel:
\beq
\hat{\sigma}_{1}&=&-{\hat{u}\over \hat{s}^2}-{1\over N^2}{2\over \hat{s}}-\frac{1} {N}\frac{1} {\hat{s}}+{1\over N^3}{1\over \hat{t}},\nonumber\\
\hat{\sigma}_{2}&=&{\hat{t}\over \hat{s}^2}+{1\over N^2}
{\hat{s}-2\hat{t}\over \hat{s}^2}+{1\over N}{\hat{u}\over 2\hat{s}\hat{t}}
+{1\over N^3}{4\hat{t}+\hat{u}\over 2\hat{s}\hat{t}}, \nonumber\\
\hat{\sigma}_{3}&=&{1\over N^2}{2\hat{s}+\hat{u}\over \hat{s}^2}-{1\over N}{1\over \hat{t}}+{1\over N^3}{1\over \hat{s}}.
\eeq
(vi) $\bar{q}q\to gg$ channel:
\beq \label{e:sigma6}
\hat{\sigma}_{1}&=&C_F{2(\hat{t}^3-\hat{u}^3)\over
\hat{s}^2\hat{t}\hat{u}}-{1\over N}{\hat{t}-\hat{u}\over \hat{s}^2}, \nonumber\\
\hat{\sigma}_{2}
&=&-C_F{2(\hat{t}-\hat{u})(\hat{s}^2+\hat{t}\hat{u})\over \hat{s}^2\hat{t}\hat{u}}
+{C^2_F\over N}{2(\hat{t}-\hat{u})\over \hat{t}\hat{u}}+
{1\over N}{\hat{t}-\hat{u}\over \hat{s}^2}, \nonumber\\
\hat{\sigma}_{3}&=&
C_F{2(\hat{t}^2-\hat{t}\hat{u}-\hat{u}^2)\over \hat{s}\hat{t}\hat{u}}
-{C^2_F\over N}{4\over \hat{t}}+{1\over N}{\hat{t}-\hat{u}\over \hat{s}^2}.
\eeq
For the charge conjugated channels (where an antiquark comes from the longitudinally polarized proton) we find $\hat{\sigma}_{\bar{a}\bar{b}\to\bar{c}\bar{d}} = \hat{\sigma}_{ab\to cd}$, where $\hat{\sigma}_{ab\to cd}$ are given in Eqs.~(\ref{e:sigma1})--(\ref{e:sigma6}). As shown in Sec.~2, there are various twist-3 distributions which 
are not independent of each other. 
In particular,
$h_L(x')$, $\tilde{h}_L(x')$, and $H_{DL}(x'_1,x')$ can be expressed in terms of
$H_{FL}(x'_1,x')$ and the transversity distribution $h_1(x')$, and thus are
``auxiliary" twist-3 distributions.\footnote{We refer the reader to Ref.~\cite{Kanazawa:2015ajw} for an extensive work on relations between twist-3 functions (including fragmentation ones) and their importance in showing the Lorentz invariance of twist-3 cross sections.} However, 
the simple structure of the
partonic cross section for $H_{FL}(x'_1,x')$ 
allows us to rewrite the cross section
in terms of $h_{1}(x')$, $h_L(x')$, and $\tilde{h}_L(x')$, as shown in Eq.~(\ref{final}), 
for the LO twist-3 cross section.  
We recall a similar simplification also occurred for the third term in
Eq.~(\ref{twist3})~\cite{Koike:2015yza}.


\section{Summary}

In this paper we have derived the twist-3 contribution from the longitudinally polarized nucleon to $A_{LT}$ in $p^\uparrow \vec{p}\to h\,X$.  
Along with the other two twist-3 pieces derived in the literature~\cite{Metz:2012fq,Koike:2015yza}, 
we now have the complete LO cross section for this process at twist-3.  
Like in the case of the twist-3 fragmentation contribution for $A_{LT}$~\cite{Koike:2015yza},
we found that the twist-3 part for the
longitudinally polarized proton can be also expressed in a simple form using one-variable quark distributions. This
will be useful for phenomenological analyses. Given that $A_{LT}$ probes different yet equally important aspects of hadronic structure as $A_N$, and the fact that RHIC has never run an experiment for this asymmetry despite being the only accelerator in the world with polarized proton beams and having measured every other proton spin configuration, we plan to conduct such a numerical study in future work.

\section*{Acknowledgments}

This work has been supported by the Grant-in-Aid for
Scientific Research from the Japanese Society of Promotion of Science
under Contract No.~26287040 (Y.K.), the RIKEN BNL
Research Center (D.P.), and in part by the NSFC under Grant No.~11575070 (S.Y.).

\end{document}